\documentclass[journal]{IEEEtran}
\usepackage{float}
\usepackage{array,multirow}
\usepackage{graphicx}
\usepackage{tabularx}
\usepackage{tabulary}
\usepackage{latexsym}
\usepackage{textgreek}
\usepackage{layout}
\usepackage[hidelinks]{hyperref} 
\usepackage{rotating}
\usepackage{amsmath,amsfonts,amssymb}
\usepackage{bm}
\usepackage[nodisplayskipstretch]{setspace}
\usepackage{dirtytalk}
\usepackage{euscript}
\usepackage{algorithmic}
\usepackage{amssymb, amsmath}
\usepackage[numbers, square, comma, sort&compress]{natbib}
\usepackage[ruled,lined]{algorithm2e}
\usepackage{epstopdf}
\usepackage[flushleft]{threeparttable}
\usepackage{gensymb}
\usepackage{subcaption}
\usepackage{siunitx}
\usepackage{url}
\usepackage{xcolor}
\usepackage{ragged2e}
\usepackage[english]{babel}
\usepackage{multicol}
\usepackage[official]{eurosym}
\usepackage{color}
\usepackage{acronym}
\usepackage{mathtools}
\usepackage[verbose]{wrapfig}
\usepackage{multirow}
\usepackage[normalem]{ulem}
\usepackage [autostyle, english = american]{csquotes}
\MakeOuterQuote{"}
\usepackage{caption} 
   \usepackage[belowskip=2pt, aboveskip=-13pt] {caption}
\newcommand{\eq}[1]{(\ref{#1})}

\acrodef{API}[API]{Application Programmable Interfaces}
\acrodef{ARPU}[ARPU]{Average Revenue per User}
\acrodef{BS}[BS]{Base Station}
\acrodef{CNN}[CNN]{Convolutional Neural Network}
\acrodef{CPU}[CPU]{Central Processing Unit}
\acrodef{DRC-RS}[DRC-RS]{Dynamic Resource Controller for Remote/Rural Sites}
\acrodef{EB}[EB]{Energy Buffer}
\acrodef{EH}[EH]{Energy Harvesting}
\acrodef{ES}[ES]{Energy Saving}
\acrodef{EM}[EM]{Energy Manager}
\acrodef{GENM}[GENM]{Green-based Edge Network Management}
\acrodef{EPC}[EPC]{Evolved Packet Core}
\acrodef{ETSI}[ETSI]{European Telecommunications Standards Institute}
\acrodef{GP}[GP]{Geometric Programming}
\acrodef{ITS}[ITS] {Intelligent Transport System}
\acrodef{LOC}[LOC]{User Location Services} 
\acrodef{LLC}[LLC]{Limited Lookahead Control}
\acrodef{LS}[LS]{Location Service}
\acrodef{LSTM}[LSTM]{Long Short-Term Memory}
\acrodef{MEC}[MEC]{Multi-access Edge Computing}
\acrodef{ML}[ML]{Machine Learning}
\acrodef{MN}[MN]{Mobile Network}
\acrodef{TIM}[TIM]{Telecom Italia Mobile}

\acrodef{NEF}[NEF]{Network Exposure Function}
\acrodef{NOES}[NOES]{NO Energy Saving}
\acrodef{NFV}[NFV]{Network Function Virtualization}
\acrodef{NIC}[NIC]{Network Interface Card}
\acrodef{QoS}[QoS]{Quality of Service}
\acrodef{RNN}[RNN]{Recurrent Neural Network}
\acrodef{RAN}[RAN]{Radio Access Network}
\acrodef{RMSE}[RMSE]{Root Mean Square Error}
\acrodef{RNN}[RNN]{Recurrent Neural Network}
\acrodef{SDN}[SDN]{Software Defined Networking}
\acrodef{UE}[UE]{User Equipment}
\acrodef{VM}[VM] {Virtual Machine}
\acrodef{VNF}[VNF]{Virtualized Network Function}

\begin{document}

\title{Remote and Rural Connectivity: Infrastructure and Resource Sharing Principles}

\author{\IEEEauthorblockN {Thembelihle Dlamini\IEEEauthorrefmark{1},  Sifiso Vilakati \IEEEauthorrefmark{2}}\\
	\IEEEauthorblockA {\IEEEauthorrefmark{1}Department of Electrical and Electronic Engineering, University of Eswatini, Kwaluseni, Eswatini}\\
	\IEEEauthorblockA {\IEEEauthorrefmark{2}Department of Statistics and Demography, University of Eswatini, Kwaluseni, Eswatini}\\
	\{tldlamini, svilakati\}@uniswa.sz
	 \vspace{-0.4cm}
	 

}

\maketitle
\thispagestyle{plain}
\pagestyle{plain}


\begin{abstract}
As Mobile Networks (MNs) are advancing towards meeting mobile users’ requirements, the \mbox{rural-urban} divide still remains a major challenge. While areas within the urban space (metropolitan mobile space) are being developed, i.e., small Base Stations (BSs) empowered with computing capabilities are deployed to improve the delivery of user requirements, rural areas are left behind. Due to challenges of low population density, low income, difficult terrain, non-existent infrastructure, lack of power grid, remote areas have low digital penetration. This situation makes remote areas less attractive towards investments and to operate connectivity networks, thus failing to achieve universal access to the Internet. In addressing this issue, this paper proposes a new BS deployment and resource management method for remote and rural areas. Here, two MN operators share their resources towards the procurement and
deployment of green \mbox{energy-powered} BSs equipped with computing capabilities. Then, the network infrastructure is shared between the mobile operators, with the main goal of enabling energy-efficient infrastructure sharing, i.e., BS and its co-located computing platform. Using this resource management strategy in rural communication sites guarantees a Quality of Service (QoS) comparable to that of urban communication sites. The performance evaluation conducted through simulations validates our analysis as the prediction variations observed shows greater accuracy between the harvested energy and the traffic load. Also, the energy savings decrease as the number of mobile users (50 users in our case) connected to the remote site increases. Lastly, the proposed algorithm achieves $51 \%$ energy savings when compared with the $43 \%$ obtained by our benchmark algorithm. The proposed method demonstrates superior performance over the benchmark algorithm as it uses foresighted optimization where the harvested energy and the expected load are predicted over a given \mbox{short-term} horizon. 
\end{abstract}


\IEEEpeerreviewmaketitle

\section{Introduction}
The evolution of the mobile and wireless communication networks into the fifth generation (5G) will play a significant role in improving the global economy. With the internet of things (IoT) dictating the way in which people communicate through information sharing and knowledge dissemination, internet coverage neesd to be improved. The capacity to provide radio coverage over a wide geographic area is a \mbox{pre-requisite} towards meeting the \mbox{ultra-low} latency requirements demanded by mobile subscribers~\cite{ericssonreport}\cite{energymanagershow}. Through the installation of a \acp{BS} and the development of the mobile and wireless communications, continuous communications can be achieved. This constitutes a gigantic step towards solving the rural/remote connectivity problem since electricity might be unreliable and it is very costly to extend grid connection to remote areas. Therefore, the provisioning of communication services in remote areas entails the use of renewable energy. Using renewable energy, coupled with sustainable energy storage solutions is a promising solution towards resolving the remote area energy predicament. \\
\indent Despite the use of green energy as a potential solution, many rural and remote areas in developed or undeveloped countries around the world are facing the challenge of unreliable \mbox{high-quality} Internet connectivity~\cite{remote6g}. This is because \ac{MN} operators are still skeptical towards making information \& communications technology (ICT) infrastructure investments in remote areas - hence the digital divide. One of the essential reasons is low expected revenue, calculated as \ac{ARPU}, which reduces companies' willingness to invest in these areas. However, with the current trends in battery and solar module costs showing a decrease, MN operators might be motivated to make investments in remote and rural areas and deploy connectivity networks. Moreover, the advent of open, programmable, and virtualized $5$G networks, will enable MN operators to overcome the limitations presented by the current \acp{MN}~\cite{energymanagershow}\cite{open5G} and make the ease of deploying open and programmable \acp{MN} a possibility. \\
\indent To extend network coverage to remote/rural areas, the use of terrestrial or \mbox{non-terrestrial} networks is proposed in~\cite{antenna_for}. In parallel, Sparse Terrestrial Networks (STN) using high towers and large antenna arrays are being developed to deliver very long transmission ranges. Here, the systems are equipped with the latest emerging antenna technologies and designs such as reconfigurable phased/inflatable/fractal antennas realized with metasurface material. Towards this, the works of~\cite{antenna_for}study the feasibility of providing connectivity to sparse areas utilizing \mbox{massive-MIMO} where the existing infrastructure of TV towers was used. In that work, it is observed that higher frequencies provide larger area coverage, provided that the antenna array area is the same. Another strategy for achieving good coverage as well as high capacity in remote/rural areas is to utilize two frequency bands, one low band and one high band, in an aggregated configuration. Following this strategy, the authors of~\cite{5g_rural_nr} combine the New Radio (NR) $3.5$ GHz and LTE $800$ MHz on a GSM grid. In addition, along the lines of long range systems, the NR is expected to support high data rates with low average network energy consumption through its lean design and massive MIMO utilization. Also, the authors of~\cite{deep_rural} extend rural coverage with STNs. Here, the large cells are created by using \mbox{long-range} links between \acp{BS} and \ac{UE}, where the long range is achieved by high towers combined with large antenna arrays and efficient antenna techniques creating narrow beams with high gain with a \mbox{line-of-sight} (LoS) or \mbox{near-LoS} connection to the UE.\\
\indent In order to end this digital divide, MNs have to \mbox{re-look} the way in which they are operating and make the necessary adjustments. One workable solution is making use of the softwarization technologies such as: \ac{SDN}, \ac{NFV}, \ac{MEC}, to be enablers for \textit{resource sharing} and \textit{edgefication}~\cite{open5G}\cite{online_pimrc}. Furthermore, the emergence of network slicing further avails new market opportunities~\cite{interdigital} for \acp{MN} to explore. In network slicing, the BS site infrastructure (\textit{resource blocks, bandwidth, computing resources}) can be shared {\it fairly} by two or more mobile operators in \mbox{real-time}. This is to effectively maximize the use of existing network resources while simultaneously minimizing the operational costs in remote sites. Also, the open and accessible shared infrastructure can enable more MN operators and Internet service providers to expand their footprint into \mbox{low-income} areas, increasing the availability of connectivity in these areas and contributing to bridging the digital divide. For continuous operation in the rural/remote communication sites, the BS empowered with computing capabilities can be \mbox{co-located} with \ac{EH} systems for harvesting energy from the environment, storing it in \acp{EB} (storage devices), and then powering the site.\\
\indent There are several forms of infrastructure sharing cases already in existence~\cite{mobilesharing}, such as the \mbox{roaming-based} sharing where the MN operators share the cell coverage for a prenegotiated time period. For example, using this \mbox{roaming-based} sharing, a \ac{UE} can employ the roaming procedure in order to connect to a foreign network. In these \say{classical} forms of sharing generally one MN operator still retains ownership of the mobile network. 
Under shared infrastructure, new entrants no longer need to incur the \mbox{often-significant} upfront cost of building their own infrastructure and can save time and resources that would otherwise be dedicated to administrative authorization and licensing. However, potential risks to
competition, governance, and implementation need to be managed to achieve the greatest benefit from infrastructure sharing.
In this article, the BS infrastructure sharing and its \mbox{co-located} computing platform (\ac{MEC} server) is done only for handling \mbox{delay-sensitive} workloads in remote/rural areas. Here, MN operators still have control of the delay-tolerant workloads to their remote clouds. This entails bringing the notion of \mbox{\it co-ownership} of the communication sites in remote/rural areas, within the \ac{MEC} paradigm, in which \textit{two} MN operators pull together their capital expenditure in order to share the deployed infrastructure, thus saving precious (already limited) economic resources for other types of expenses. Then, in order to effectively manage the BS sites deployed in remote/rural areas, procedures for dynamic network control (\textit{managing network resources when MN operators share fairly their network infrastructure}) and agile management are required. This will assist in efficiently delivering a comparable \ac{QoS} in remote/rural areas to that of urban areas.\\
\indent The work done in this article is an extension of~\cite{online_pimrc}, where \ac{BS} sleep modes and \ac{VM} \mbox{soft-scaling} procedures were employed towards energy saving in remote sites. In \cite{online_pimrc}, energy savings were obtained through \mbox{short-term} traffic load and harvested energy predictions, along with energy management procedures. However, the considered energy cost model does not take the caching process, tuning of transmission drivers, and the use of \mbox{container-based} virtualization into account. In addition, the considered communication site belongs to \textit{one} MN operator, i.e., the site infrastructure was not shared between multiple operators. Therefore, the \mbox{computing-plus-communication} energy cost model is the main motivation for this article, where the BS site is shared among multiple operators in order to handle \mbox{delay-sensitive} workloads only. 
One application of our model (strategy) corresponds to the current situation that has been caused by the new coronavirus (COVID-19) pandemic. The pandemic has reshaped our living preferences such that rural (remote) areas are now becoming more and more attractive. This can motivate MN operators to deploy networks in such areas and then share their communication infrastructure and the computing resources that are \mbox{co-located}. The contributions of this article are summarized as follows: 
\begin{itemize}
\item [1)] A \ac{BS} empowered with computing capabilities \mbox{co-located} with a \ac{EH} system is considered, whereby the MN operators share the BS site infrastructure (i.e., \textit{bandwidth, computing resources}) for handling \mbox{delay-sensitive} workloads within a remote/rural area.
\item [2)] In order to enable foresighted optimization, a \mbox{short-term} future communications site workload and harvested energy is forecasted using a \ac{LSTM} neural network~\cite{lstmlearn}. 
\item [3)] An online \mbox{controller-based} algorithm {\it called} \ac{DRC-RS} for handling infrastructure sharing and managing the communication site located in remote/rural areas is developed. The proposed algorithm is based on the \ac{LLC} approach and resource allocation procedures with the objective of enabling for infrastructure sharing (BS and its \mbox{co-located} computing platform) and resource management within remote and rural communication sites.
\item [4)] \mbox{Real-world} harvested energy and traffic load traces are used to evaluate the performance of the proposed optimization strategy. The numerical results obtained through simulation show that the proposed optimization strategy is able to efficiently manage the remote/rural site and also allows the sharing of the network infrastructure.
\end{itemize}
\begin{table*} [h!]
	\caption{Comparison with existing works.}
	\label{tab_opt1}
	\begin{threeparttable}
	\center
	\begin{tabular} {|l|l|l|l|l|}
		\hline 
		{\bf Feature} & {\bf Edge computing}  & {\bf Method Used} & {\bf Forecasting} & {\bf Objective}\\ 
		\hline
		RAN sharing~\cite{sharingRAN} & No & Linear programming & No & Max. QoS\\ \hline
		Traffic load exploitation~\cite{gamebasedsharing} & No & Game theory & No & Min. spending cost\\ \hline
		Contractual backup~\cite{strategicsharing} & No & Contract design under & No & Max. resource utilization\\
		& & symmetric information & & and profits\\ \hline
	    Multiple-seller single-buyer~\cite{sanguanpuak} & No & Stochastic geometry & No & Cost minimization\\ 
	    & & & & Guarantee of QoS \\ \hline
	    Communication and  & Yes & \ac{LSTM} & Yes & Min. energy consumption\\
	    Computation [\textbf{Proposed}] & & \ac{LLC} & & Guarantee of QoS\\
		\hline 
	\end{tabular}
	\begin{tablenotes}
      \small
      \item Yes: considered; No: not considered
    \end{tablenotes}
    \end{threeparttable}
\end{table*}
In order to achieve these, the remainder of this article is organized as follows: Section~\ref{sec:rel} discusses previous research works related to the one undertaken in this article. Section~\ref{sec:sys} describes the proposed system model using detailed explanation on the operation of each network element. The mathematical problem formulation is given in Section~\ref{sec:prob} together with the details of the optimization problem and the proposed \ac{DRC-RS} online algorithm. In Section~\ref{sec:eval}, a performance evaluation of the proposed online algorithm is presented using simulation results and statistical discussions. The conclusions of this article are then given in Section~\ref{sec:concl}.
\section{Related Work}\label{sec:rel}
\noindent MN operators generally have complete ownership and control of their network and their networks are characterized by an inflexible and monolithic infrastructure. Such a rigid status quo incapacitates networks of the required versatility, hence they cannot cope with the dynamically changing requirements. As a result, in their current state, meeting the heterogeneity and variability of future MNs is an impossible task. As mobile and wireless networks evolve, MN operators are faced with the daunting task of keeping up and coping with the accelerated \mbox{roll-out} of new technologies. Due to this fast-paced technological advancements, large and frequent investments are made in order to cope with the new services and network management phases. This proactive network operation and management consequently increases the network operating costs, which reduces the intended profits. Thus, in order to reduce the \mbox{per-MN} operator investment cost, the sharing of network infrastructure between mobile operators is an attractive solution. To this effect, the authors in~\cite{sharingRAN} proposed a \ac{RAN} sharing scheme where MN operators share a single radio infrastructure while maintaining separation and full control over the backhauling and their respective core networks. In that paper, a mixed integer linear programming (MILP) formulation is proposed for determining the sharing configurations that maximize the \ac{QoS}, and a cooperative game theory concept is used to determine stable configurations as envisioned by the MN operator. The regulatory enforcement towards offering the best service level for the users and the greedy approach considered in that paper reduce the effectiveness of infrastructure sharing, as both approaches do not promote fairness among \ac{MN} operators.
In addition, the work of~\cite{gamebasedsharing} employs an infrastructure sharing algorithm towards energy savings by exploiting the under utilization of the network during \mbox{low-traffic} periods. In their work, a \mbox{game-theoretic} framework was proposed in order to enable the MN operators to individually estimate the \mbox{switching-off} probabilities that reduce their expected financial cost. Apart from the energy efficiency benefits, the proposed scheme allows the participating MN operators to minimize their spending costs independently of the strategies of the coexisting MN operators. Despite of the presented benefits, it is worth noting that infrastructure sharing should be considered for both low- and high-traffic periods, which is the focus of this paper. However, due to the existence of competition between the different MNs, collaboration in this infrastructure sharing is a primary requisite. In order to enforce such a collaboration between competitors, the authors in~\cite{strategicsharing} proposed a strategic network infrastructure sharing framework for contractual backup reservation between a small/local network operator of limited resources and uncertain demands, and one resourceful operator with potentially redundant capacity. Here, one MN operator pays for network resources reserved for use by its subscribers in another MN operator, while in turn the payee guarantees the availability of the resources. Then, in~\cite{sanguanpuak}, the problem of infrastructure sharing among MN operators is presented as a \mbox{multiple-seller} \mbox{single-buyer} business. In their contribution, each \ac{BS} is utilized by subscribers from other operators and the owner of the BS is considered as a seller of the BS infrastructure while the owners of the subscribers utilizing the BS are considered as buyers. In the presence of multiple seller MN operators, it is assumed that they compete with each other to sell their network infrastructure resources to potential buyers. \\
\indent The aforementioned works consider BS infrastructure sharing towards lowering operational cost, either by switching on/off the BSs, while maintaining the network control. In addition, infrastructure sharing is treated as a business case instead of a cooperative effort towards boosting connectivity in remote/rural areas. If one MN operator is treated as a seller while the other one as a buyer if it uses its network resources, this becomes a business venture. For example, one MN operator might be using the resource reservation technique, whereby it reserves resources for other small operators. Again, here the other party has to pay in order to use those facilities. However, it is worth mentioning that the works done in~\cite{sharingRAN}\cite{gamebasedsharing}\cite{strategicsharing}\cite{sanguanpuak} do not consider infrastructure sharing with the \ac{MEC} paradigm and the consideration of green energy has been overlooked. Those that are within the \ac{MEC} paradigm they share their \textit{own} network resources, among themselves in order to handle spatially uneven computation workloads in the network. Their objective being to avoid large computation latency at overloaded small BSs as well as to provide high quality of service (QoS) to end users. The details of how internal infrastructure sharing is conducted cannot be covered in this article, interested readers are referred to~\cite{chen2018computation}. Table~\ref{tab_opt1} above summarizes the differences of the infrastructure sharing strategy from existing works.
\section{System Model}\label{sec:sys}
\begin{figure}[h!]
	\centering
	\includegraphics[width = \columnwidth]{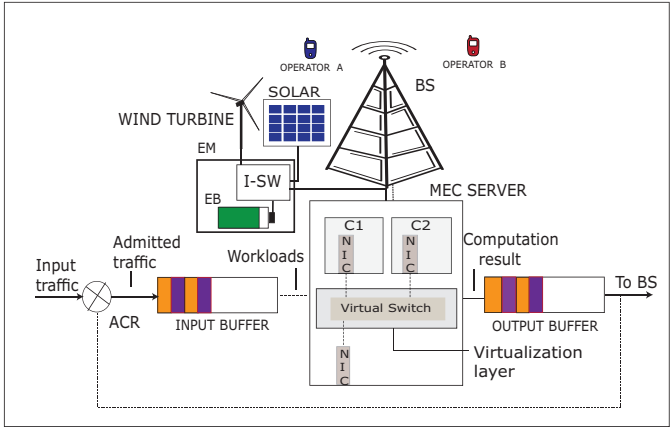}
	\caption{The remote/rural BS site infrastructure consisting of the BS co-located with the MEC server both powered by green energy obtained from solar radiation and wind turbine.}
	\label{fig:remotesite}
\end{figure}
In this paper, we consider a remote/rural site network scenario as illustrated in Fig.~\ref{fig:remotesite} above. Each network apparatus (BS, MEC server) in the figure is mainly powered by renewable energy harvested from wind and solar radiation, and it is equipped with an \ac{EB} for energy storage. The stored energy is shared by the edge server and the BS system. The \ac{EM} is an entity responsible for selecting the appropriate energy source to fulfill the \ac{EB}, and also for monitoring the energy level of the \ac{EB}.  Then, the intelligent \mbox{electro-mechanical} switch (I-SW) aggregates the energy sources to fulfill the \ac{EB} level.
\noindent The proposed model in Fig.~\ref{fig:remotesite} above is \mbox{cache-enabled}, TCP/IP offload capable (i.e., enables {\it partial} offloading in the server's \ac{NIC} such as checksum computation~\cite{sohan2010characterizing}). The virtualized MEC server, which is \mbox{co-located} with the \ac{BS}, is assumed to be hosting $C$ containers (see C1, C2 in Fig.~\ref{fig:remotesite}). Also, it has an input and output buffer for holding the workloads. It is assumed that some of the BS functions are virtualized as pointed in~\cite{BS_virtualization} as the \ac{MEC} node is composed of a virtualized access control router which acts as an access gateway for admission control. The virtualized access control router (ACR) is responsible for local and remote routing, and it is locally hosted as an application. Here, it is assumed that the remote/rural site infrastructure is shared between {\it two} MN operators through a \mbox{pre-existing} agreement, where a common microwave backhaul or a \mbox{multi-hop} wireless backhaul relaying is used for accessing remote clouds or the Internet. Moreover, a \mbox{discrete-time} model is considered, whereby the time is discretized as \mbox{$t = 1,2,\dots$} time slots of a fixed duration $\tau$.
\subsection{Input Traffic and Queue Model}
\begin{figure}[t]
	\centering
	\includegraphics[width = \columnwidth]{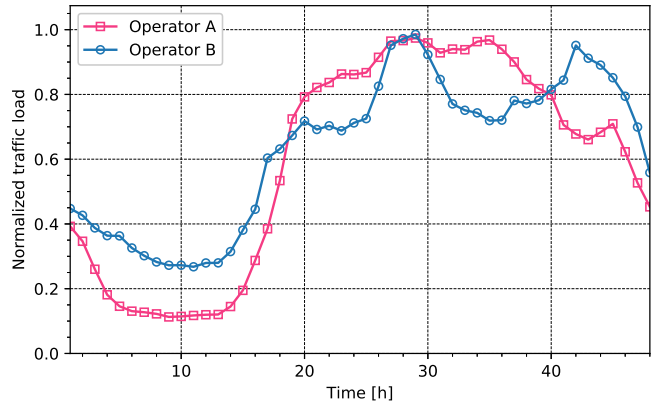}
	\caption{Normalized BS traffic loads behavior representing two MN operators represented as operator A and B.}
	\label{fig:trace_load}
\end{figure}

\noindent In the communication site, the BS is the connection point anchor and the computing platform processes the currently assigned \mbox{delay-sensitive} tasks by \mbox{self-managing} its own local virtualized storage/computing resources. Also shown in Fig.~\ref{fig:remotesite} above is an input buffer of size $L_{\rm in}$, a reconfigurable computing platform and the related switched virtual LAN, an output queue of size $L_{\rm out}$; and a controller that \mbox{re-configures} the \mbox{computing-plus-communication} resources and also performs the control of input/output traffic flows. Since the workload demand exhibits a diurnal behavior in remote/rural areas, forecasting the mobile operator's workload can help towards network infrastructure sharing. Thus, in order to emulate the remote site traffic load $L(t)$ (from $|\nu(t)|$ users), real MN traffic load traces from~\cite{bigdata2015tim} are used. It is assumed that \textit{only} Operators A and B share the remote/rural BS site, and their traffic load profiles are denoted by $L_{\rm A} (t)$ and $L_{\rm B}(t)$ ([bits]), respectively. It is also assumed that $L_{\rm A}(t)$ (or $L_{\rm B}(t)$) consists of $0.8$ \mbox{delay-sensitive} workloads $\gamma_{\rm A}(t)$ (or $\gamma_{\rm B}(t)$) and the remainder is delay-tolerant. The total admitted workload is denoted by $\gamma^*(t) = \gamma_{\rm A}(t) + \gamma_{\rm B}(t)$, i.e., $\gamma^*(t) \leq L_{\rm in}$). The input/output (I/O) queue of the system are assumed to be \mbox{loss-free} such that the time evolution of the backlogs queues follows Lindley's equations. The normalized BS traffic load behavior representation of the two mobile operators is illustrated in Fig. \ref{fig:trace_load} above.
\subsection{Communication and Computing Energy Cost Model}
\noindent For the BS system deployed in the remote/rural area, the total energy consumption $\theta_{\rm SITE}(t)$ (measured in $\SI{} {\joule}$) at time slot $t$ consists of the BS communications, denoted by $\theta_{\rm COMM}(t)$, and computing platform processes, related to computing, caching, and communication, which is denoted by $\theta_{\rm COMP}(t)$. Thus, the energy consumption model at time slot $t$ is formulated as follow, inspired by~\cite{steering}:
\begin{equation}
	\theta_{\rm SITE}(t) = \theta_{\rm COMM}(t) +  \theta_{\rm COMP}(t).     
	\label{eq:siteconsupt}
\end{equation}
\noindent The \ac{BS} energy consumption processes $\theta_{\rm COMM}(t)$ constitutes of the sum of the following: 
\begin{equation}
\theta_{\rm COMM}(t) = \sigma(t)\theta_0 + \theta_{\rm load}(t) + \theta_{\rm bk} + \theta_{\rm data}(t)\gamma^*(t)\,,
\end{equation}
\noindent where $\sigma (t)\in \{0,1\}$ is the BS switching status indicator, with $1$ representing the active mode while $0$ indicates the power saving mode. $\theta_0$ is the load independent constant value  representing the operation energy, \mbox{$\theta_{\rm load} (t) = L(t) (2^{\frac{r_0}{\zeta(t) W}}-1)N_0 (K)^\alpha \beta^{-1}$} the load dependent transmission power to the served subscribers that guarantees low latency services at a target rate $r_0$. The term $W$ is the channel bandwidth, $\zeta(t)$ is the fraction of the bandwidth used by the mobile users from operator A and B, while $\alpha$ and $\beta$ are the path loss exponent and the path loss constant, respectively. The term $K$ denotes the average distance between two BSs within the same region, and $N_0$ is the noise power. The parameter $\theta_{\rm bk}$ represents the constant microwave backhaul transmission energy cost, and $\theta_{\rm data}(t)$ (fixed value in J/byte) is the \mbox{inter-communication} cost incurred by exchanging data between the BS and MEC interfaces.\\
\indent Next, we discuss the MEC server processes that make up $\theta_{\rm COMP}(t)$. With $\gamma^*(t)$ being the currently admitted workload to be processed, let $\gamma_c(t) \leq \gamma_{\rm max}, c = 1, \dots, C(t)$, denote the size of the task that the scheduler allocates, per container, bounded by the set maximum amount $\gamma_{\rm max}$. This is such that the following constraint: $\sum_{c=1}^{C(t)} \gamma_c(t) = \gamma^*(t)$, guarantees that the overall workload is partitioned into $|C(t)|$ parallel tasks. 
This load distribution is motivated by the shares feature~\cite{migrationpower} that is inherent in virtualization technologies. This enables the resource scheduler to efficiently distribute resources amongst contending containers, thus guaranteeing the completion of the computation process within the expected time.
Thus, the set of attributes which characterize each container are: $\{\psi_c(t), \theta_{{\rm idle},c}(t), \theta_{{\rm max},c}(t), \Delta, f_c(t) \},$, where $\psi_c(t) = (f_c(t)/f_{\rm max})^2$ is the container utilization function, and $f_{\rm max}$ is the maximum available processing rate for container. Here, $f_c(t) \in [f_0, f_{\rm max}]$ denote the processing rates of container $c$, whereby the term $f_0$ is the zero speed of the container, e.g., deep sleep or shutdown. The term $\theta_{{\rm idle},c}(t)$ represents the static energy drained by the container $c$ in its idle state, $\theta_{{\rm max},c}(t)$ is the maximum energy that container $c$ can consume, and $\Delta$ is the maximum \mbox{per-slot} and \mbox{per-container} processing time ([s]).\\
\indent Within the computing platform, the energy drained due to the active containers, denoted by $\theta_{\rm CP}(t)$, is induced by the \ac{CPU} share that is allocated for the workload, and it is given by:
\begin{equation}
  \theta_{\rm CP}(t) = \sum_{c=1}^{C(t)}\theta_{{\rm idle}, c}(t) + \psi_{c}(t) (\theta_{{\rm max},c}(t)-\theta_{{\rm idle}, c}(t)).
  \label{eq:cp}
\end{equation}
It should be noted that within the edge server there is the virtualization layer with switching capabilities (see Fig.~\ref{fig:remotesite}). Thus, the processing rates are switched from the processing rates of the previous time instance ($t-1$), denoted by $f_c(t-1)$, to the present instance ($t$), denoted by $f_c(t)$. This entails an energy cost, denoted by $\theta_{\rm SW}(t)$, which is defined as:
\begin{equation}
  \theta_{\rm SW}(t) = \sum_{c =1}^{C(t)} k_e (f_c(t)-f_c(t-1))^2,
  \label{eq:sw}
\end{equation}
where $k_e$ represents the \mbox{per-container} reconfiguration cost caused by a unit-size frequency switching which is limited to a few hundreds of $\SI{}{\metre\joule}$ per $(\rm MHz)^2$. \\
\indent The MEC server can perform TCP/IP computation processing in the network adapter in order to minimize the CPU utilization. Such process incurs an energy that is drained, denoted by $\theta_{\rm OF}(t)$, which is obtained as:
\begin{equation}
   \theta_{\rm OF}(t) = \delta(t) \theta_{\rm idle}^{\rm nic}(t) + \theta_{\rm max}^{\rm nic}(t),
   \label{eq:of}
\end{equation}
where $\theta_{\rm idle}^{\rm nic}(t)$ (a non-zero value) is the energy drained by the adapter when powered but with no data transfer processes. This avails an opportunity to reduce the \mbox{non-zero} value to zero energy. For this, $\delta(t) = (0, 1)$ is the switching status indicator, with 1 indicating the active state and $0$ representing the idle state. Then, $\theta_{\rm max}^{\rm nic}(t)$ is the maximum energy drained by the network adapter process and it is obtained in a similar way as in~\cite{steering}. \\
\indent In order to keep the \mbox{intra-communication} delays at a minimum, it is assumed that each container $c$ communicates with the resource scheduler through a dedicated reliable link that operates at the transmission rate of $r_c(t)$ [(bits/s)]. Thus, the power drained by the $c^{\rm th}$ \mbox{end-to-end} connection is given by:
\begin{equation}
  P_c^{\rm net}(t) = \Psi_c (\overline{rtt_c} \, r_c(t))^2,
\end{equation}
where $c = 1, \dots, C(t), \overline{rtt_c}$ is the average \mbox{round-trip-time} of the $c^{\rm th}$ \mbox{intra-connection}, and $\Psi_c$ (measured in $\SI{}{\watt})$ is the power consumption of the connection when the product, i.e., the \mbox{round-trip-time}, which is by \mbox{communication-rate-unit-valued}. Therefore, after $\gamma_c(t)$ has been allocated to container $c$, the corresponding communication energy consumed by the $c^{\rm th}$ links is, denoted by $\theta_{\rm LK}(t)$, is obtained as:
\begin{equation}
   \theta_{\rm LK}(t) = P_c^{\rm net}(t) (\gamma_c(t)/r_c(t)) \equiv (2\Psi_c/(\tau - \Delta)) (\overline{rtt}_c \gamma_c(t))^2. 
   \label{eq:lk}
\end{equation}
\noindent In practical application scenarios, the maximum \mbox{per-slot} communication rate within the \mbox{intra-communications} is generally limited by a \mbox{pre-assigned} value $r_{\rm max}$, thus the following hard constraint must hold: $\sum_{c=1}^{C(t)} r_c(t) = \sum_{c=1}^{C(t)} (2\gamma_c(t)/ (\tau - \Delta)) \leq r_{\rm max}$. We also note that there exists a \mbox{two-way} per task execution delay where each link delay is denoted by $\varrho_c(t) = \gamma_c(t)/r_c(t)$. In this work, we assume that the overall delay equates to $2\,\varrho_{c}(t) + \Delta$.\\ 
\indent To dequeue the computational results from the output buffer, denoted by $L_{\rm out}$, the optical tunable drivers are used for the data transfers processes. A \mbox{trade-off} between the transmission speed and the number of active drivers per time instance is required to reduce the energy consumption. For data transfers, $|D(t)| \leq D$ drivers are required  for transferring $l_d(t) \in L_{\rm out}$. The energy drained by the data transfer process, denoted by $\theta_{\rm LS}(t)$, consists of the energy for utilizing each fast tunable driver, denoted by $m_d(t) [(J/s)]$ (a constant value), the target transmission rate $r_0$, and $L_{\rm out}$. Thus, the energy is obtained as follows:
\begin{equation}
  \theta_{\rm LS}(t) = \sum_{d=1}^{D(t)} (m_d(t) l_d(t))/{r_0},
  \label{eq:ls}
\end{equation}
where the parameters are obtained similar to~\cite{steering}. \\
\indent To minimize the network traffic from the remote/rural site to the remote clouds, some of the frequently requested internet content are cached locally, more especially viral contents. The caching process contribute to the energy consumption within the site, denoted by $\theta_{\rm CH}(t)$, and it is obtained as~\cite{steering}: 
\begin{equation}
  \theta_{\rm CH}(t) = \overline{\lambda} (t)\,(\theta_{\rm TR} (t) + \theta_{\rm CACHE}(t)),
  \label{eq:cache}
\end{equation}
where $\theta_{\rm TR} (t)$ represents the power consumption due to transmission processes, $\theta_{\rm CACHE}(t)$ is the power consumption contributed by the caching process with its \mbox{intra-communication}, and $\overline{\lambda} (t)$ is the response time function for viral content~\cite{large_youtube}. \\
\indent In overall, the resulting \mbox{communication-plus-computing} processes incurs an energy cost $\theta_{\rm COMP}(t)$, per slot $t$, which is given by Eqs.~(\ref{eq:cp}), (\ref{eq:sw}), (\ref{eq:of}), (\ref{eq:lk}), (\ref{eq:ls}), (\ref{eq:cache}), as follows:
\begin{equation}
    \begin{aligned}
    \theta_{\rm COMP}(t) & =  \theta_{\rm CP}(t) + \theta_{\rm SW}(t) + \theta_{\rm OF}(t) \\
                           & + \theta_{\rm LK}(t) + \theta_{\rm LS}(t) + \theta_{\rm CH}(t).
   \end{aligned}
   \label{eq:mec_cost}
\end{equation}
\subsection{Energy Harvesting and Demand Profiles}
\begin{figure}[t]
	\centering
	\includegraphics[width = \columnwidth]{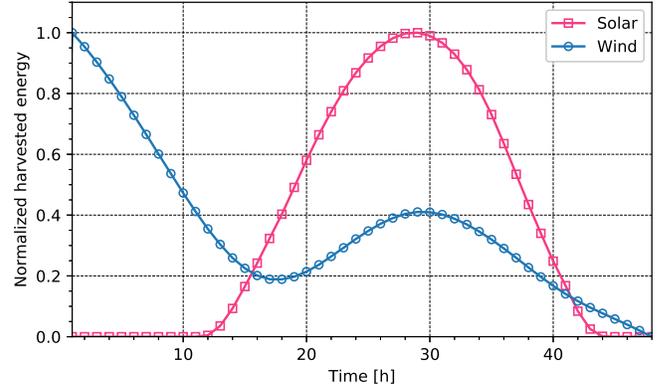}
	\caption{Example traces for harvested solar traces and wind traces from~\cite{belgium}.}
	\label{fig:energy_trace}
\end{figure}

\noindent The rechargebale energy storage device is characterized by its finite energy storage capacity $E_{\rm max}$, and the energy level reports are periodically pushed to the \mbox{DRC-RS} application in the \ac{MEC} server. In this case, the \ac{EB} level $B(t)$ is known, which enables for the provisioning of the required communication and computing resources in the form of the required containers, transmission drivers, and the transmission power in the BS. To emulate the profiles, the amount of harvested energy $H(t)$ in time slot $t$ is obtained from \mbox{open-source} solar and wind traces from a farm located in Belgium~\cite{belgium}, and they are as shown in Fig.~\ref{fig:energy_trace}.
\noindent The data in the dataset matches the time slot duration of ($\SI{30} {\minute}$) used in this work and it is the result of daily environmental records. 
In this work, the wind energy is selected as a power source during the solar energy \mbox{off-peak} periods. The available \ac{EB} level $B(t + 1)$ located at the offgrid site evolves according to the following dynamics: 
\begin{equation}
   \mbox{$E(t + 1) = \min\{E(t) + H(t) - \theta_{\rm SITE}(t)- a(t), E_{\rm max}\}$},
\label{eq:offgrid}
\end{equation}
where $E (t)$ is the energy level in the battery at the beginning of time slot $t$, $\theta_{\rm SITE}(t)$ represents the site energy consumption, {\it see} Eq.~\eq{eq:siteconsupt} above, and $a(t)$ is leakage energy. However, it is worth noting that the energy level $E(t)$ is updated at the beginning of time slot $t$, whereas $H(t)$ and $\theta_{\rm SITE}(t)$ are  only known at the end of $t$. Thus, the energy constraint at the off-grid site must be satisfied for every time slot: $\theta_{\rm SITE}(t) \leq E(t)$. Therefore, for decision making, the online controller simply compares the received EB level reports with two \mbox{set-points} ($0 < E_{\rm low} < E_{\rm up} < E_{\rm max}$), the lower $E_{\rm low}$ and upper $E_{\rm up}$ energy thresholds. Here, $E_{\rm low}$ is the lowest EB level that the off-grid site should reach and $E_{\rm up}$ corresponds to the desired energy buffer level at the site. If $E(t) < E_{\rm low}$, then the site is said to  be energy deficient, and a suitable energy source at each time slot $t$ is selected on the forecast expectations, i.e., the expected harvested energy $\hat{H}(t)$.
\section{Problem Formulation}
\label{sec:prob}
\noindent In this section, the optimization problem is formulated to obtain an energy efficient infrastructure sharing and resource management procedures through \mbox{short-term} traffic load and harvested energy forecasting. The overall goal is to enable energy efficient infrastructure sharing and resource management, within remote and rural communication sites, and in turn guaranteeing a comparable \ac{QoS} to that of urban areas, with reduced energy consumption in remote/rural sites. 
\subsection{Optimization Problem}
\label{opt}
\noindent Within the BS, the allocated bandwidth $W$ is shared between mobile subscribers from operator A and B, and within the computing platform, the containers (i.e., as the computing resources) and the underlying physical resources (e.g., \ac{CPU}) are shared among the users who offloaded their \mbox{delay-sensitive} workloads.
To address the aforementioned problem, two cost functions are defined, namely, F1 and F2, where (F1) is defined as: $\theta_{\rm SITE}(t)$ (F1), weighs the energy drained in the BS site due to transmission and computing processes; and (F2) which accounts for the comparable \ac{QoS} is defined as: $(\gamma^*(t) - L_{\rm in})^2$. Regarding this formulation, it is worth noting that F1 tends to push the system towards \mbox{self-sustainability} solutions and F2 favors solutions where the delay sensitive load is entirely admitted in the computing platform by the router application, taking into account the expected energy to be harvested. The corresponding (weighted) cost function is defined as:
\begin{equation}
\label{eq:Jfunc_2}
\begin{aligned}
J(\zeta, \sigma,C,D, t) & \stackrel{\Delta}{=} \Upsilon \, \theta_{\rm SITE}(\zeta(t), \sigma(t), C(t), D(t), t)\\
                   & + \overline{\Upsilon}(\gamma^*(t) - L_{\rm in}(t))^2 \, ,
 \end{aligned}
\end{equation}
where $\Upsilon = [0,1]$ is the weight used to balance the two functions, and $\overline{\Upsilon} \stackrel{\Delta}{=} 1 - \Upsilon$. Hence, starting from the current ti,e slot $t = 1$ to the finite horizon $T$, the time is discretized as follows: $t = 1,2, \dots, T$), thus the optimization problem is formulated as follows: 
\begin{eqnarray}
        \label{eq:objt_2}
        \textbf{P1} & : & \min_{\mathcal{N}} \sum_{t=1}^T J(\zeta, \sigma, C,D, t)  \\
        && \hspace{-1.25cm}\mbox{subject to:} \nonumber \\
        {\rm A1} & : & \sigma(t) \in \{0,1\}, \nonumber \\
        {\rm A2} & : & \beta \leq C(t) \leq C, \nonumber \\
        {\rm A3} & : & E(t) \geq E_{\rm low} , \nonumber \\ 
        {\rm A4} & : & 0 \leq \gamma_{c}(t) \leq \gamma_{\rm max}, \nonumber \\
        {\rm A5} & : & 0 \leq f_{c}(t) \leq f_{\rm max}, \nonumber \\
        {\rm A6} & : & \mbox{$r_{\rm min} \leq r_c(t) \leq r_{\rm max}$}, \nonumber\\
        {\rm A7} & : & \mbox{$\theta_{\rm SITE}(t) \leq E(t)$}, \nonumber\\
        {\rm A8} & : & \max \{2\,\varrho_{c}(t)\} + \Delta = \tau_{\rm max}, \quad t=1,\dots, T \, , \nonumber     
\end{eqnarray} 
where the set of objective variables to be configured at slot $t$ in the BS system and MEC server is defined as \mbox{$\mathcal{N} \stackrel{\Delta}{=} \{\zeta(t), \sigma(t), C(t), \{\psi_c(t)\}, \{P_c^{\rm net}(t)\}, \{\gamma_c(t)\}, \delta(t), D(t)\}$}. These settings handle the transmission and computing activities using the following constraints. Here, Constraint A1 specifies the BS operation status (either {\it power saving} or {\it active}), A2 forces the required number of containers, $C(t)$, to be always greater than or equal to a minimum number \mbox{$\beta \geq 1$}. The purpose of this is to be always able to handle mission critical communications. The constraint A3 ensures that the \ac{EB} level is always above or equal to a preset threshold $E_{\rm low}$, to guarantee {\it energy \mbox{self-sustainability}} over time. Furthermore, A4 bound the maximum workloads of each running container $c$, with $c = 1,\dots, C(t)$, A5 represents a \mbox{hard-limit} on the corresponding \mbox{per-slot} and \mbox{per-VM} processing time. A6 forces $r_c(t)$ to fall in a desired range: [$r_{\rm min}, r_{\rm max}$] of transmission rates and A7 ensures that the  energy consumption at the site is bounded by the available energy in the EB. A8 offers the hard \ac{QoS} guarantees within the computing platform. From \textbf{P1}, it is noted that there exists a \mbox{non-convex} component $P_c^{\rm net}(t)$, from $\theta_{\rm LK}(t)$. In this case, the Geometric programming (GP) concept can be used to convert $\theta_{\rm LK}(t)$ into a convex function similar to~\cite{steering}. Thus, in order to solve {\bf P1} in~\eq{eq:objt_2}, the \ac{LLC} approach~\cite{hayes_2004}, GP technique, and heuristics, is used towards obtaining the feasible system control inputs $\eta (t) = (\zeta(t), \sigma(t), C(t), \{\psi_c(t)\}, \{P_c^{\rm net}(t)\}, \{\gamma_c(t)\}, \delta(t), D(t))$ for $t=1,\dots,T$. Well, it should be noted that~\eq{eq:objt_2} can iteratively be solved at any time slot $t \geq 1$, by just redefining the time horizon as $t^\prime = t, t+1, \dots, t+T-1$.
\subsubsection{Feasibility and QoS guarantees}
Regarding the feasibility of the problem, the following formal results holds.\\
\noindent\textbf{Proposition 1.} Feasibility conditions\\
\indent \textit{The following two inequalities:}
\begin{equation}
   (r_{\rm max}/2)(\tau - \Delta) \geq L_{\rm in}
\end{equation}
\begin{equation}
    \sum_{c=1}^{C(t)} f_c(t) \Delta \geq r_{\rm min}
\end{equation}
\textit{guarantees that the infrastructure sharing and resource reconfiguration problem is feasible}. \qquad\qquad\qquad\qquad\qquad\qquad $\square$

Since the reported conditions assure that P1 admits the solution, we then consider the corresponding QoS properties. In this regard, it is safe to point out that A6 and A8 lead to the following hard bounds on the resulting \mbox{communication-plus-computing} delay.\\
\noindent\textbf{Proposition 2.} Hard \ac{QoS} guarantees\\
\indent\textit{Firstly, the feasibility conditions of Proposition 1 must be met. Next, we let random variables measure the following: the random queue delay of the input queue $\tau_{IQ}$, the service time of the input queue $\tau_{SI}$, the queue delay of the output queue $\tau_{OQ}$, and the service time of the output queue $\tau_{SO}$. Thus, the following QoS guarantees hold:} 
\textit{the random total delay ($\tau_{\rm tot} \stackrel{\Delta}{=} \tau_{IQ} + \tau_{SI} + \tau_{OQ} + \tau_{SO}$) induced by the computing platform is limited (in a hard way) up to:}
\begin{equation}
\tau_{\rm tot} \leq ((L_{\rm in} + L_{\rm out})/ r_{\rm min}) + 2.
\end{equation}
Thus, the reported QoS guarantee lead to the conclusion that the remote/rural site can handle \mbox{delay-sensitive} workloads while meeting the bound in A8.
\subsection{Infrastructure Sharing and Resource Allocation}
\label{infra}
\noindent In this subsection, the predictions for the BS traffic load and energy consumption, the description of the remote/rural site system dynamics, and the proposed online \mbox{controller-based} algorithm are presented. 
\subsubsection{Prediction of exogenous processes}
\label{predic}
\noindent Two exogenous processes are considered in this work: the harvested energy $H(t)$ and the BS traffic loads $L(t)$. In order to generate the predictions ($\hat{H}(t), \hat{L}(t)$), the \ac{LSTM} neural networks~\cite{lstmlearn} were adopted. Thus, the \mbox{LSTM-based} predictor has been trained to give an output of the the forecasts for the required number of future time slots $T$. The trained LSTM network consists of an input layer, a single hidden layer consisting of $40$ neurons, for $80$ epochs, for a batch size of $4$; and an output layer. For training and testing purposes, the dataset was split as $70\%$ for training and $30\%$ for testing. As for the performance measure of the model, the \ac{RMSE} is used.
\subsubsection{Remote/Rural site system dynamics}
\label{rurdynamics}
\noindent In order to effectively manage the remote/rural site, an adaptive implementation of the controller is developed. Its purpose is to compute the solutions of both the infrastructure sharing and resource configurations \mbox{on-the-fly}. For this purpose, an online \mbox{controller-based} algorithm is proposed and is outlined in {\bf Algorithm \ref{tab:genm}} below.
\begin{small}
\begin{algorithm}[h!]
\begin{tabular}{l l}
{\bf Input:}  & $s(t)$ (current state) \\
{\bf Output:} & $\eta^{*}(t)$  (control input vector)\\
01:		& \hspace{-1cm} Parameter initialization\\
		& \hspace{-1cm} ${\mathcal G}(t) = \{s(t)\}$ \\
02:		& \hspace{-1cm} {\bf for} ($k$ within the prediction horizon of depth $T$) {\bf do}\\
		& \hspace{-1cm}\quad - $\hat{L}(t+k)$:= forecast the workload  \\
		&\hspace{-1cm}\quad - $\hat{H}(t+k)$:= forecast the energy\\
		& \hspace{-1cm}\quad - ${\mathcal G}(t+k) = \emptyset$ \\
03:		& \hspace{-1cm}\quad {\bf for} (each $s(t)$ in ${\mathcal G}(t+k)$) {\bf do}\\
             & \hspace{-1cm}\qquad - generate all reachable states $\hat{s}(t+k)$\\
             & \hspace{-1cm}\qquad - ${\mathcal G}(t+k) = {\mathcal G}(t+k) \cup \{\hat{s}(t+k)\}$\\
04:		& \hspace{-1.1cm} \quad\quad {\bf for} (each $\hat{s}(t+k)$ in $\mathcal G(t+k)$) {\bf do}\\
            & \hspace{-1.1cm}\qquad\quad - calculate the corresponding $\theta_{\rm SITE}(\hat{s}(t+k))$\\
            & \hspace{-1.1cm}\qquad\quad taking into account of $\zeta(t)$, and $l_d(t)$ from $L_{\rm out}(t)$\\
		& \hspace{-1.1cm} \quad\quad {\bf end for}\\
		& \hspace{-1.1cm}\quad\quad {\bf end for}\\
		& \hspace{-1cm} \quad {\bf end for}\\
05:		& \hspace{-1cm} - obtain a sequence of reachable states yielding\\
        & \hspace{-1cm}\quad the best system input\\	
06:		& \hspace{-1cm} {$\eta^{*}(t):=$ control leading from $s(t)$ to $\hat{s}_{\min}$}\\
07:		& \hspace{-1cm} {\bf Return $\eta^{*}(t)$}
\end{tabular}
\caption{DRC-RS Algorithm Pseudocode}
\label{tab:genm}
\end{algorithm}
\end{small}

\noindent At this point, it should be noted that at time slot $t$ the system state vector is $s(t) = (\zeta(t), \sigma(t), C(t), D(t), E(t))$ and the applied input vector that drivers the system towards the desired behaviour. These drivers perform bandwidth sharing, adaptive BS power transmission, autoscaling and reconfiguration of containers, and tuning of the optical drivers and is denoted by $\eta^*(t) = \{\zeta(t), \sigma(t), C(t), \{\psi_c(t)\}, \{P_c^{\rm net}(t)\}, \{\gamma_c(t)\}, \delta(t), D(t)\}$. The system behavior is described by the \mbox{discrete-time} \mbox{state-space} equation, adopting the \ac{LLC} principles~\cite{hayes_2004}:
\begin{equation}
     s(t + 1) = \Phi(s(t), \eta(t)) \, , 
\end{equation}
where  $\Phi(\cdot)$ is a behavioral model that captures the relationship between $(s(t),\eta(t))$, and the next state $s(t + 1)$. This relationship accounts for the amount of energy drained $\theta_{\rm SITE}(t)$, that harvested $H(t)$, which together lead to the next buffer level $E(t+1)$ through Eq.~\eq{eq:offgrid}. The \ac{DRC-RS} algorithm, finds the best control action vector $\eta^*(t)$ that yields the desired system behaviour within the remote/rural site. Note that $P_c^{\rm net}(t)$ is obtained using the CVXOPT toolbox and $\gamma_c(t), C(t),$ is obtained following the procedure outlined in remark 1 in~\cite{steering}. The entire process is repeated every time slot $t$ when the controller can adjust the behavior given the new state information. The state values of $s(t)$ and $\eta(t)$ are measured and applied at the beginning of the time slot $t$, whereas the offered load $L(t)$ and the harvested energy $H(t)$ are accumulated during the time slot and their value becomes known only at the end of it. This means that, being at the beginning of time slot $t$, the system state at the next time slot $t+1$ can only be estimated, which is formally written as:
\begin{equation}
       \hat{s}(t + 1) = \Phi(s(t),\eta(t)) \,.
       \label{eq:state_forecast}
\end{equation}
At this regard, it is worth noting that the control actions are taken after exploring only a limited prediction horizon, yielding a limited number of possible operating states. In order to ensure system stability, we rely on the notion that a system is said to be stable under control, if for any state, it is always possible to find a control input that forces it closer to the desired state or within a specified neighborhood of it~\cite{llcprediction}.
\subsubsection{Dynamic Resource Controller for Remote/Rural Sites}
\label{alg}
\noindent The edge network management algorithm pseudocode is outlined in Algorithm 1 above and it is based on the LLC principles, where the controller obtains the best control action $\eta^*(t)$. Starting from the {\it initial state}, the controller constructs, in a \mbox{breadth-first} fashion, a tree comprising all possible future states up to the prediction depth $T$. The algorithm proceeds as follows: \\
\begin{table} [t]
	\caption{System Parameters.}
	\center
	\begin{tabular} {|l| l|l|}
		\hline 
		{\bf Parameter} & {\bf Value} \\ 
		\hline
		Microwave backhaul power, $\theta_{\rm bk}$ & $\SI{50}{\watt}$\\
		BS operating power $\theta_0$, & $\SI{10.6}{\watt}$\\
		Max. number of containers, $C$ &  $20$\\
		Min. number of containers, $\beta$ & $1$ \\
		Time slot duration, $\tau$ &  $\SI{30} {\minute}$\\
		Container $c$ (idle state), $\theta_{{\rm idle}_c}(t)$ & $\SI{4} {\joule}$\\
		Container $c$ (max), $\theta_{{\rm max},c}(t)$ & $\SI{10} {\joule}$\\
		Reconfiguration cost, $k_e$ & $ 0.005 \rm J/(\rm MHz)^2$\\
		NIC in idle state, $\theta_{\rm idle}^{\rm nic}(t)$ & $13.1 \rm J$\\
		Max. allowed processing time, $\Delta$ & $\SI{0.8} {\second}$\\
		Processing rate set, $\{f_c(t)\}$  & $\{0,50,70,90,105\}$\\
		Bandwidth, $W$ & $1 {\rm MHz}$\\
		Max. allocated $c$ workload $\gamma_{\rm max}$ & 10 MB\\
		Max. number of drivers, $D$ & $6$\\
		Noise spectral density, $N_0$ & $-174 \, {\rm dBm/Hz}$\\
		Max. container $c$ load, $\gamma_{\max}$ & $ 10$ MB\\
		Driver energy, $m_d(t)$ & $1 \, \rm J/s$\\
		Target transmission rate, $r_0$ & $1 \, \rm Mbps$\\
		Leakage energy, $a (t)$ & $2\, \mu \rm J$\\
		Energy storage capacity, $E_{\rm max}$ & $\SI{490} {\kilo\joule}$\\
		Lower energy threshold, $E_{\rm low}$  & $30$\% of $E_{\rm max}$\\
		Upper energy threshold, $E_{\rm up}$  & $70$\% of $E_{\rm max}$
\\
		\hline 
	\end{tabular}
	\label{tab_opt}
\end{table}
\indent A search set $\mathcal G$ consisting of the current system state is initialized (line 01), and it is accumulated as the algorithm traverse through the tree (line 03), accounting for predictions, accumulated workloads at the output buffer, past outputs and controls, operating intervals. The set of states reached at every prediction depth $t+k$ is referred to as $\mathcal G(t+k)$ (line 02). Given $s(t)$, the traffic load $\hat{L}(t+k)$ and harvested energy $\hat{H}(t+k)$ is estimated first (line 02), and generate the next set of reachable control actions by applying the accepted workload $\gamma^{*}(t+k)$, energy harvested and shared bandwidth fraction $\zeta (t+k)$. The cost function corresponding to each generated state $\hat{s}(t+k)$ is then computed (line 04), where $\hat{s}(t+k)$ take into account of $l_d$ as observed from $L_{\rm out}(t)$. Once the prediction horizon is explored, a sequence of reachable states yielding minimum energy consumption is obtained (line 05). The control action $\eta^{*}(t)$ corresponding to $\hat{s}(t+k)$ (the first state in this sequence) is provided as input to the system while the rest are discarded (line 06). The process is repeated at the beginning of each time slot $t$.
\section{Performance Evaluation}
\label{sec:eval}
\noindent In this section, some selected numerical results for the scenario of Section~\ref{sec:sys} are shown. The parameters that were used in the simulations are listed in Table~\ref{tab_opt} above.
\begin{figure}[t]
	\centering
	\begin{subfigure}[t]{\columnwidth}
		\centering
		\includegraphics[width = \columnwidth]{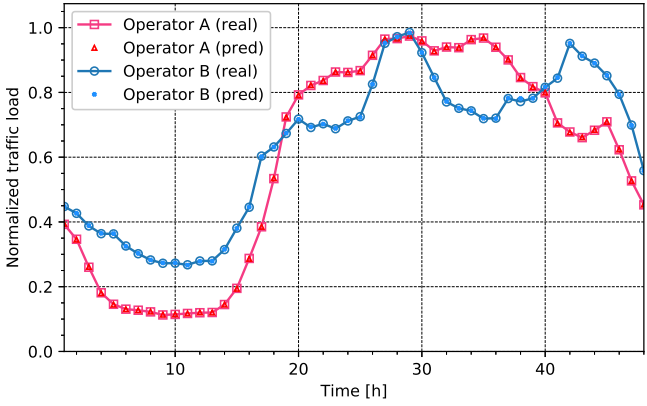}
		\caption{One-step ahead predictive mean value for $L(t)$.}
		\label{fig:bs_load}	
	\end{subfigure}
	\quad
	\begin{subfigure}[t]{\columnwidth}
		\centering
		\includegraphics[width = \columnwidth]{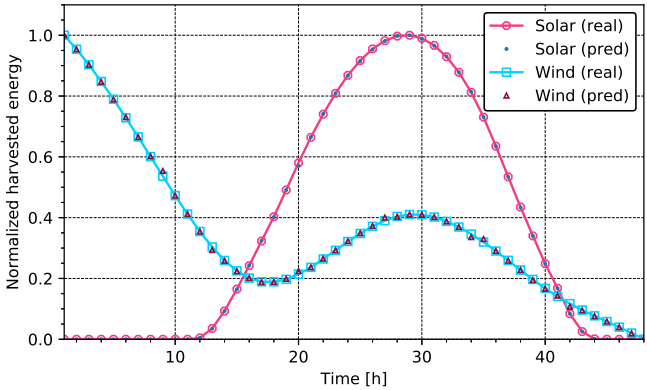}
		\caption{One-step ahead predictive mean value for $H(t)$.}
		\label{fig:energy_load}	
	\end{subfigure}
	\centering
	\caption{One-step online forecasting for both $L(t)$ and $H(t)$ patterns.}
	\label{fig:patterns}
\end{figure}
\subsection{Simulation setup}
A BS empowered with computation capabilities deployed in a rural/remote area is considered in this setup. Our time slot duration $\tau$ is set to $\SI{30} {\minute}$ and the time horizon is set to $T = 3$ time slots. For simulation, Python is used as the programming language.
\subsection{Numerical results}
\textit{Data preparation:} The information from the used mobile and energy traces is aggregated to the set time slot duration. The mobile traces are aggregated from $\SI{10}{\minute}$ observation time to $\tau$. As for the wind and solar traces, they were aggregated from $\SI{15}{\minute}$ observation time to $\tau$. The used datasets are readily available in a public repository (\textit{see}~\cite{traces}).\\
In Fig.~\ref{fig:patterns}, the real and predicted values for traffic load from operator A and B, harvested energy is shown. Here, the forecasting routing tracks each value and predict it over \mbox{one-step}. The shown selected prediction results are for operator A and B, Solar, and Wind. Then, Table~\ref{tab:pred} below shows the the average \ac{RMSE} of the normalized harvested energy and traffic load processes ($L_A, L_B$), for different time horizon values, $T \in \{1,2,3\}$. In the table, the term $H_{\rm wind} (t)$ represent the forecasted values for energy harvested from wind turbines and $H_{\rm solar} (t)$ is for the harvested energy from solar panels. From the obtained results, the prediction variations are observed between $H(t)$ and $L(t)$ when comparing the average RMSE. The measured accuracy is deemed good enough for the proposed optimization.
\begin{table}[H]
\footnotesize
\centering
\caption{Average prediction error (RMSE) for harvested energy and
traffic load processes, both normalized in [0,1].}
\begin{tabulary}{1.0\textwidth}{|L|L|L|L|}
\hline
  & {$T = 1$} & {$T = 2$} & {$T = 3$} \\
\hline
$L_A (t)$ & 0.070 & 0.090 & 0.011\\ \hline
$L_B (t)$ & 0.050 & 0.070 & 0.010\\ \hline
$H_{\rm wind}(t)$  & 0.011 & 0.013 & 0.016\\ \hline
$H_{\rm solar}(t)$ & 0.050 & 0.070 & 0.090\\
\hline
\end{tabulary}
\label{tab:pred}
\end{table}
The \mbox{DRC-RS} algorithm is benchmarked with another one, named Resource Reservation Manager (RRM), which is inspired by the backup reservation agreement from~\cite{strategicsharing}. In the RRM, the network resources are reserved per time slot based on a \mbox{set-point} threshold percentage. Both algorithms make use of the learned information.\\
\begin{figure}[h]
	\centering
	\includegraphics[width = \columnwidth]{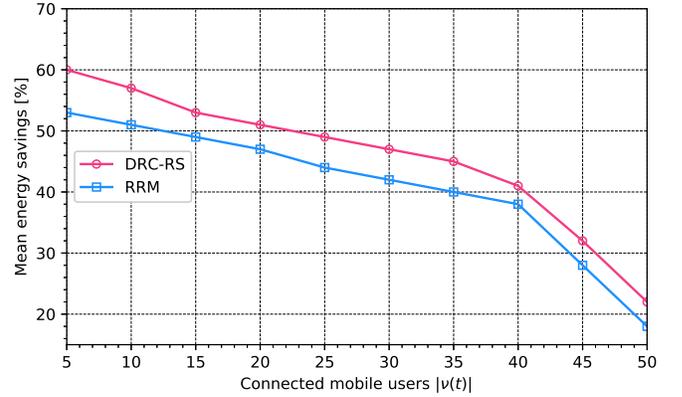}
	\caption{Energy savings versus number of users connected to the BS.}
	\label{fig:bsusers}
\end{figure} 
Figure~\ref{fig:bsusers}, shows the average energy savings obtained within the offgrid system. Here, the number of users connected to the remote site is increased from $|\nu(t)|$ = $5$ to $50$, using an incremental step size of $5$. The obtained energy savings are with respect to the case where the BS site is dimensioned for maximum expected capacity (maximum value of $\theta_{\rm COMM}(t), \theta_{\rm COMP}(t)$). From the results, as expected, it is observed that the energy savings decrease as the number of mobile users connected to the remote site increases. The \mbox{DRC-RS} outperforms the RRM algorithm. At this regard, we note that the communication site will accept users as long as energy harvesting projections are positive.\\
\begin{figure}[t]
	\centering
	\includegraphics[width = \columnwidth]{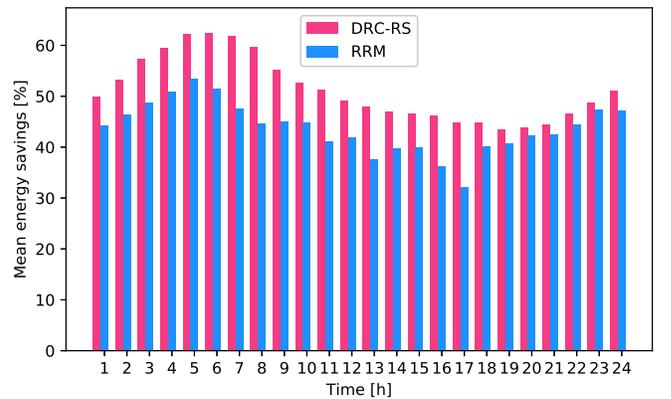}
	\caption{Mean energy savings for the remote/rural site system.}
	\label{fig:rmsite}
\end{figure} 
Then, Fig.~\ref{fig:rmsite} shows the average energy savings for the edge system. Here, the BS group size is set to $|\nu(t)| = 20$ and the obtained energy savings results are with respect to the case where no energy management procedures are applied, i.e., the BS is dimensioned for maximum expected capacity (maximum value of $\theta_{\rm SITE} (t)$, $\forall t$) and the MEC server provisions the computing resources for maximum expected computation workload (maximum value of $\theta_{\rm MEC} (t)$, with $C = 20\,  \text{containers}, \forall t$). The average results of \mbox{DRC-RS} ($k_e = 0.05, \gamma_{\rm max} = 10$ MB) show energy savings of $51 \%$, while RRM  achieves $43 \%$ on average. The effectiveness of the BS management procedure, autoscaling and reconfiguration of the computing resources, and on/off switching of the fast tunable laser drivers, coupled with foresighted optimization is observed in the obtained numerical results.
\section{Conclusions}
\label{sec:concl}
The challenge of providing connectivity to remote/rural areas will be one of the pillars for future mobile networks. To address this issue, in this paper, we present an infrastructure sharing and resource management mechanism for handling \mbox{delay-sensitive} workloads within a remote/rural site.
Numerical results, obtained with \mbox{real-world} energy and traffic load traces, demonstrate that the proposed algorithm achieves mean energy savings of $51 \%$ when compared with the $43 \%$ obtained by our benchmark algorithm. Also, the energy that can be saved decreases as the number of user connected to the BS increases, with a guarantee of serving more users as long the green energy is available.
The energy saving results are obtained with respect to the case where no energy management techniques are applied in the remote site.
\section*{Data Availability}
In this paper, open-source datasets for the mobile network (MN) traffic load, solar, and wind energy have been used. The details are as follows: (1) the real MN traffic load traces used to support the findings of this study were obtained  from the Big Data Challenge organized by Telecom Italia Mobile (TIM) and the data repository has been cited in this article. (2) The real solar and wind traces used to support the findings of this study have also been cited in this article.

\bibliographystyle{IEEEtran}
\scriptsize
\bibliography{biblio_new}
\end{document}